\\

Title: On the concept of "efficiency loophole" and rejection of local realism

Author: M. R. Mahdavi

Comments: 7 pages, 2 figure. For any comments contact <<mr_mahdavi@hotmail.com>>

Subj-class: quant-ph; physics

\\

In this paper I shall demonstrate that an analysis of the concept of "efficiency loophole" leads to the conclusion that any type of charge carried by an elementary particle must be distributed on the surface of a non-spherical object. As a result I shall show that; for example, the electrostatic field of an electron as seen by an observer, must be in the form of quantized superluminal pulses. Therefore, "efficiency loophole" cannot be used in support of local realistic models. In view of the available experimental results we conclude that a local realistic interpretation of quantum mechanics has now been empirically refuted. However, it can be shown that the inferred superluminal signals do not necessarily contradict the facts of Lorentz transformations and only introduce a weak non-locality into the laws of nature. For this reason we make a preliminary investigation about a possible connection between quantum mechanics and non-spherical charge distributions. We conclude that this idea deserves to be further investigated and its validity or otherwise can be verified experimentally.

\\

Aspect's [1] and similar Bell-type experiments carried out since 1982 seemingly provide overwhelming support in favour of the Copenhagen interpretation of quantum mechanics [2]. However, the proponents of local realistic models have managed to challenge the interpretations of the experimental results, partly by questioning the validity of the basic assumptions and processing of the experimental data [3], and partly by advancing the concept of "efficiency loophole" [4,5]. In a recent work [6] Larsson has shown that by utilizing this concept a local realistic model can mimic the quantum probabilities, and that a sort of unification between the Copenhagen and Bohm interpretations of quantum mechanics can also be achieved.

The concept of "efficiency loophole" is based on the proposition that aside from random errors the detection efficiency is also a manifestation of a hidden physical property of the particles being detected. Thus, for example, a particle may arrive at the detector but because of the hidden property it may not be detected. The result of a measurement, say, on the spin of a particle will now yield not two, but one of the three results, $\pm \hbar/2$ and no results.

A fundamental examination, however, will show that "efficiency loophole" is a problematic concept that immediately leads to problems and contradictions that can in no way be resolved within the framework of present day physics. This was to be expected, since a hidden variable theory presupposes an underlying physics of whose particles, fields, or laws of motion we are totally ignorant. In the present note we shall not consider any of the serious problems that ensue from the above concept; rather, we wish to show that, in any event, it cannot be used in advocating a local realistic model, and shall briefly consider its efficacy in regards to a non-local realistic interpretation of quantum mechanics.

Consider a detector specifically designed to register the arrival of an electron at a point in space by measuring its electrostatic field (we could use a gravity or a magnetic detector but the conclusions would be the same). If because of the hidden physical property the arrival of the electron is not registered it then follows that for a time period all electrostatic interactions between the electron and the constituent particles of the detector were terminated. This "no interaction" situation can arise in the following way.

The "no interaction" situation can arise only if the electrostatic field of the electron is essentially non-isotropic, or in another word, if the electronic charge is not uniformly distributed on the surface of a spherical object. However, since all evidence point to an isotropic electrostatic field for the electron, one will now be forced to further speculate that: it must be as a result of the internal motion (motions) of the electron, as evidenced by its spin, that on the *average* one observes an isotropic electrostatic field. Let us now consider the implications of this possibility. For the purpose of the present discussion it will not be necessary to speculate on the form of the charge distribution. Therefore, we will merely assume that a certain amount of charge $-q_e$ is distributed on a portion $\beta A$ of a sphere whose surface area is equal to A. Therefore, if the electron lacked any type of internal motion then, one could only view its electrostatic field from direction lying inside a solid angle $d\Omega = 2\pi\beta$. Clearly, in order that a non-spherical charge distribution may have a significant affect on the outcome of experiments the value of $\beta$ must be much less than one.

Now suppose that we wish to measure the electrostatic field of an electron with a fine instrument whose response time is less than a certain amount $\Delta t_a$. Because of the internal motion of the electron its electrostatic field will come within the view of the instrument for a time period $\Delta t_a$ and a field intensity $E = (1+\beta^{-1})E_a$ will be registered ( as schematically shown in **Fig.1**). After this the electrostatic field of the electron will be lost from the view of the instrument for a time period $\Delta t' = (1+\beta^{-1})\Delta t_a$ until such time as when the spin of the electron once again brings its electrostatic field within the view of the instrument. Clearly, $E_a$ the average field intensity as measured by a coarse instrument, as normally is the case, will be equal to $E_a = E/(1+\beta^{-1})$.

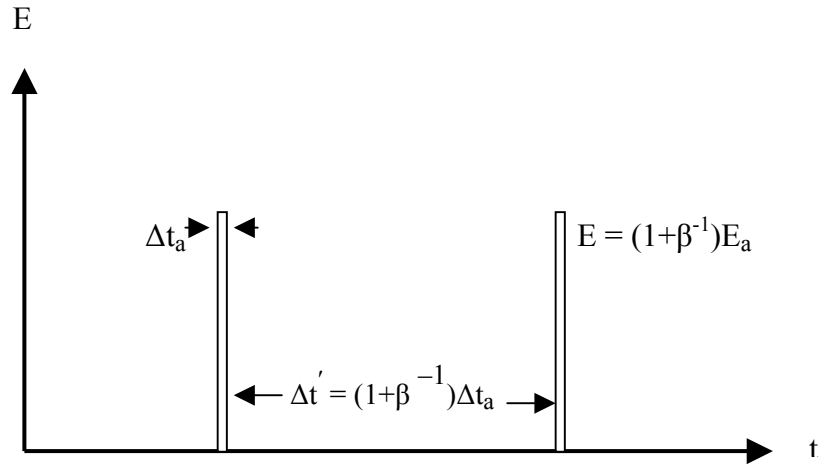

**Fig.1**

Next, consider that in the classical conception of things the electrostatic field of a charge is in the form of continuous lines issuing radially from it and that the speed of travel of electrostatic signals through space has to be assumed to be equal to that of light. Now, our analysis has shown that, what in classical physics is considered as a continuous field line, in reality (if "efficiency loophole" is a valid concept) is a train of electric pulses traveling through space. Clearly, a train of such pulses would imitate the classical field and Lorentz transformations would retain their form- only as time-averaged transformations- if the speed of travel of each pulse were equal to $C = (1+\beta^{-1})c$. Thus, when we observe the electrostatic field of an electron we first receive a pulse that travels with the speed of $C = (1+\beta^{-1})c$. However, the next pulse, as it were, lingers on for a time period $\Delta t'= (1+\beta^{-1})\Delta t_a$ before starting its travel through space. Hence the average speed comes out as being equal to $c$.

Now it will be seen that so long as an electron moves with a uniform velocity we cannot be aware of the existence of the superluminal signals. However, when an electron is subjected to an external field, its state, for example its spin, will be altered. The information about the altered state of the electron will now be conveyed to a distant point via the agency of the superluminal signals. Thus, our analysis leads to the conclusion that "efficiency loophole" cannot be advanced in supporting a local realistic interpretation of quantum mechanics. The question now arises: "should we dismiss the idea of "efficiency loophole" and the inferred non-spherical charge distribution altogether?" In order to answer this question we carry out the following preliminary analysis[1].

As forerunners to part of our arguments we mention some ideas that have already appeared in connection with stochastic theories of quantum mechanics. In the Bohm theory [7,8] it is assumed that a particle undergoes some sort of Brownian motion. This motion is assumed to arise as a result of fluctuations in the ψ field. Thus for example, a segment of a linear motion of an electron is in fact the average result of a number of random motions. In a paper on the stochastic origin of the Schrödinger's equation, Davidson [9] attributes the peculiar and non-classical features of quantum mechanics as being inherent in a statistical description of the radiative reaction force. In that paper Davidson also mentions the possibility that the ~2.7 K background microwave radiation may be the cause of the statistical variations in radiative reaction force. Since the Bohm theory yields the same results as ordinary quantum mechanics we may argue that the assumption about the Brownian motion of an elementary particle is a correct one, and adopting Davidson's suggestion we may say that there is observational evidence in support of Bohm's idea.

Now imagine an instant of time when an electron is moving along a straight line and we expect to see its electrostatic field shortly after. If as a result of interaction with the background radiation the electron changes its course of motion, then the coming into view of its electrostatic field might be delayed (equally it might be advanced). Thus, there is always a probability that a sequence of such random changes in the direction of motion of the electron may result in an inordinate lengthening of the time period $\Delta t' = (1+\beta^{-1})\Delta t_a$ during which time the electron is lost from ones view. As a concrete example of this effect consider an electron, having insufficient energy to climb out of a potential well. Despite the adverse condition there is still a probability that the combination of the internal motion of the electron plus a sequence of non-classical trajectories will allow the electron to move along a trajectory, along which it fails to "see" the electric potential. As a result it may now tunnel out of the potential well.

Next, consider that if we were to observe an electron over macroscopic distances, then at those instances that our instrument registers the position of the electron, the registering arises as a result of physical interactions. In this situation we have the particle aspect of the electron. On the other hand, in the example just cited, quantum tunneling arose because for some period of time electrostatic interactions between the elementary particles were terminated. In this situation the particle aspect is missing from the picture, and the de Broglie waves, whatever their nature, determine the course of motion of the particle.

---

[1]. The idea of a non-spherical charge distribution will be taken up in a subsequent paper where we will support it with almost conclusive evidence.

If there is any truth to the above conclusion we may now advance the following general principle. A stable quantum system, *e. g.*, an atom or a nucleus, can be obtained only if all physical interactions between the constituent particles of the system are terminated. Thus, all the particles of the quantum system are essentially free and the problem of acceleration and emission of electromagnetic radiation no longer arise.

According to our analysis there is an inherent indeterminacy, consequent upon the very physical make-up of elementary particles, which precludes the possibility of carrying out precise measurement. In certain situations, as in the example of quantum tunneling, the time period of this inherent indeterminacy is inordinately increased. In such situations the description of a particle or a quantum system can only be given in terms of a probability theory, and strictly speaking, one cannot beforehand predict the precise outcome of an experiment. The state of an elementary particle or quantum system is revealed only when a measurement has been carried out and *the measurement has yielded a definite result*.

Also, according to the above analysis an inherent indeterminacy is the root-cause of quantum mechanics. Hence, we should not be surprised if, as in the case of Larsson's work [6], the inclusion of an inherent indeterminacy into a local realistic model would yield results, mimicking the quantum mechanical probabilities.

In conclusion we may say that the concept of "efficiency loophole" fails in its intended purpose. Consequently, in view of the results obtained in Bell-type experiments we may say that local realism has been empirically refuted. However, seeing that the superluminal signals do not contradict the facts of Lorentz transformations and that the implied non-spherical charge distribution may allow us to understand quantum effects in terms of an inherent indeterminacy, we propose that it is a concept that deserves to be investigated further. A search for superluminal electrostatic signals may well be the quickest way, not only for settling the issue of the interpretation of quantum mechanics, but also as the first practical step towards the development of sub-quantum theory. However, before embarking on such an experiment, the experimenter would like to know at least the approximate value of the speed of the superluminal pulses. In regards to this we make the following observations.

The factor $\alpha^{-1} = (1+\beta^{-1})$ is dimensionless and from the preceding discussions it is apparent that there is no reason for it to appear in the equations of physics at the macroscopic level. On the other hand this is a factor that is representative of a physical attribute of elementary particles and is a measure of the intermittent interactions between particles at the quantum level. Intuitively, we may expect $\alpha^{-1}$ to make its appearance in equations at the quantum level. Now, we indeed do have a dimensionless constant, namely, $\alpha$ the fine-structure constant, which makes its first appearance in the equations of physics at the atomic level. On this basis we tentatively suggest that the speed of travel of superluminal signals through space is $C \sim 137c$ and suggest the following experiment as a way of verifying the validity or otherwise of the idea of non-spherical charge distribution.

An electron beam is chopped so as to produce an electron pulse moving with velocity $\mathbf{v}_x$ in the X-direction (**Fig. 2**). The electron pulse passes the point A; midway between two metallic balls that at a later time will constitute the poles of an electric dipole. When the electron pulse reaches the point B a voltage V is impressed between the two balls. Now, assume that electrostatic signals travel, as imagined in the classical case, with the velocity $c$. We place the collimator at a point C so that the time for the electrostatic field to travel from A to C will be greater than the time required for the electron pulse to travel from B to C. In this case the electron pulse will not be subjected to the dipole field. Therefore, it will not suffer a deflection and after passing through the collimator will arrive at the detector.

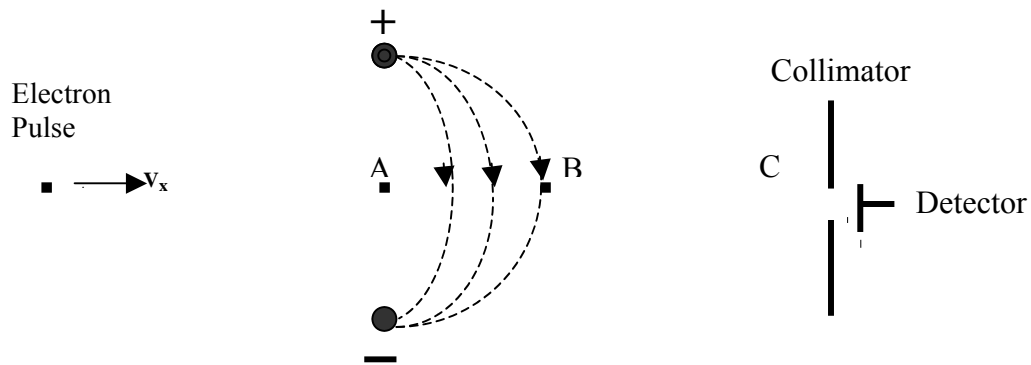

**Fig. 2**

On the other hand if the electrostatic field is in the form of superluminal pulses then, with suitable values for **v**, V, and d (the size of the orifice in the collimator), the electron pulse will suffer a deflection and will not arrive at the detector.

In principle this set-up can be used for measuring the velocity of the superluminal pulses by increasing the distance AB until the point is reached where the electron pulse eventually arrives at the detector. The speed of the superluminal pulses is then obtained from the relation

$$C = (AB/BC) \cdot c \qquad (3)$$